\documentclass[%
 reprint,
 superscriptaddress,
 amsmath,amssymb,
 aps,]{revtex4-2}

\usepackage{graphicx}
\usepackage{dcolumn}
\usepackage{bm}
\usepackage[retainorgcmds]{IEEEtrantools}
\usepackage{subfig}
\usepackage[dvipsnames]{xcolor}

\usepackage[normalem]{ulem}
\usepackage{siunitx}
\usepackage{braket}
\usepackage{nicefrac}
\usepackage{natbib}
\usepackage{soul}

\renewcommand{\vec}[1]{\boldsymbol{\mathrm{#1}}}
\newcommand{\unVec}[1]{\hat{e}_{#1}}

\newcommand{\bk}{\vec{k}}
\newcommand{\lk}{\vec{l}_{n \bk}}
\newcommand{\sk}{\vec{s}_{n \bk}}

\begin{document}

\preprint{APS/123-QED}

\title{Spin and orbital Edelstein effect in a bilayer system with Rashba interaction}

\author{Sergio Leiva M.}
 \email{sergio-tomas.leiva-montecinos@physik.uni-halle.de}
\affiliation{Institut für Physik, Martin-Luther-Universität Halle-Wittenberg, D-06099 Halle (Saale), Germany}
\author{Jürgen Henk}%
\affiliation{Institut für Physik, Martin-Luther-Universität Halle-Wittenberg, D-06099 Halle (Saale), Germany}
\author{Ingrid Mertig}%
\affiliation{Institut für Physik, Martin-Luther-Universität Halle-Wittenberg, D-06099 Halle (Saale), Germany}
\author{Annika Johansson}%
\affiliation{Max Planck Institute of Microstructure Physics, Halle, Germany}

\date{\today}

\begin{abstract}
The spin Edelstein effect has proven to be a promising phenomenon to generate spin polarization from a charge current in systems without inversion symmetry. In recent years, a current-induced orbital magnetization, called orbital Edelstein effect, has been predicted for various systems with broken inversion symmetry, using the atom-centered approximation and the modern theory of orbital magnetization. In this work, we study the current-induced spin and orbital magnetization for a bilayer system with Rashba interaction, using the modern theory of orbital magnetization and Boltzmann transport theory in relaxation time approximation. We find that the spin Edelstein effect is significantly larger than the orbital contribution. Furthermore, the orbital Edelstein response can be enhanced, suppressed, and even reversed, depending on the relation of the effective Rashba parameters of each layer. A sign change of the orbital polarization is related to an interchange of the corresponding layer localization of the states. 
\end{abstract}

\maketitle


\section{\label{sec:Introduction}Introduction}

Effective spin-charge interconversion is crucial for the realization of novel spintronic devices~\cite{wolf2001spintronics, soumyanarayanan2016emergent, manchon2017spin}. One prominent and intensely studied effect providing charge-to-spin conversion is the (spin) Edelstein effect (EE)~\cite{aronov1989nuclear, edelstein1990spin, inoue2003diffuse, kato2004current, gambardella2011current}, also known as Aronov-Lyanda-Geller-Edelstein effect~\cite{johansson2016theoretical}, inverse spin-galvanic effect~\cite{Ganichev2002, gambardella2011current}, or current-induced spin polarization. In systems with broken inversion symmetry, such as surfaces, interfaces, or systems lacking inversion symmetry in their crystal structure, the application of an external electric field, or a charge current, generates a homogeneous spin polarization due to spin-orbit coupling. Similarly, an injected spin current induces a net charge current in these systems via the Onsager reciprocal of the EE, the inverse Edelstein effect (IEE)~\cite{Shen2014}. The importance of the EE and the IEE for spintronics is due to the ability to create and control spin currents and spin polarization in a non-magnetic material solely by an applied charge current and vice versa. 

The first and most common systems for which the EE has been predicted are two-dimensional (2D) systems with Rashba spin-orbit coupling \cite{rashba1960properties,bychkov1984oscillatory,bychkov1984properties}, where the spin polarization typically arises in-plane and perpendicular to the current direction~\cite{edelstein1990spin}. The EE has been found to occur also in Weyl semimetals \cite{johansson2018edelstein, yang2021chiral}, chiral materials \cite{Shalygin2012, Furukawa2017, Furukawa2021, Calavalle2022, Suzuki2023, Tenzin2023Collinear}, oxide interfaces \cite{varotto2022direct, johansson2021spin}, topological insulators \cite{Yokoyama2010, yokoyama2011current, Luo2016, Zhang2016}, transition metal dichalcogenides (TMDs) \cite{Cysne2021nano, Lee2022, InglaAyns2022, Cysne2023}, noncentrosymmetric superconductors \cite{Chirolli2022}, and other quantum materials \cite{Han2018}. 

Besides the spin, electrons can also carry an orbital moment, which can give rise to a finite net magnetization. In most ferromagnets, the spin contribution to the equilibrium magnetization is dominant, and the orbital contribution is negligible~\cite{meyer1961experimental, ceresoli2010first, lopez2012wannier}. However, analogously to nonqequilibrium spin transport effects, orbital transport can occur~\cite{Vitale2018, Go2018, Canonico2020, Cysne2021TMDs, Lee2021, Pezo2022, Cysne2022, Costa2023, yoda2015current, yoda2018orbital, go2017toward, salemi2019orbitally, johansson2021spin, varotto2022direct, busch2023orbital}. Thus, the orbital Edelstein effect (OEE) corresponds to a current-induced orbital density, or current-induced orbital magnetization, in systems with broken inversion symmetry~\cite{levitov1985magnetoelectric, yoda2015current, yoda2018orbital, go2017toward, salemi2019orbitally, johansson2021spin, varotto2022direct}. In contrast to equilibrium ferromagnetism, the OEE has been found to be comparable or even larger than the SEE~\cite{salemi2019orbitally, johansson2021spin, varotto2022direct}. However, since the position operator is not well-defined in translationally invariant systems , the calculation of the orbital magnetization (OM) is not trivial in periodic systems~\cite{thonhauser2011theory}. In order to avoid this problem, the angular momentum operator is evaluated in disjunct spheres around the atoms. This standard method, known as the atomic-centered approximation (ACA), provides accurate and computationally efficient results for some materials, while for others, where the nonlocal contributions are relevant, the ACA approximation fails to accurately estimate the OM \cite{thonhauser2011theory, hanke2016role}. A more precise and complete alternative, including the nonlocal contributions, is the so-called modern theory of orbital magnetization \cite{chang1996berry,sundaram1999wave_packet, ceresoli2006orbital, xiao2005berry}, proposed for translationally invariant materials \cite{hanke2016role}.

The modern theory of OM has been implemented in several density-functional theory codes \cite{Nikolaev2014, ceresoli2010first, lopez2012wannier} and tight-binding models \cite{yoda2015current,yoda2018orbital}, primarily to study bulk ferromagnetic materials and heterostructures. However, the need for translational invariance of the modern theory implies a problem for interfaces and, generally, two-dimensional (2D) systems. The modern theory has recently been extended to treat the OEE in polar metals, insulator surfaces, and semi-infinite systems \cite{hara2020current}.

In this work, we apply the modern theory of OM to a two-dimensional electron gas (2DEG) modeled by an effective Rashba Hamiltonian in a bilayer system with spin-independent interlayer hopping, following the formalism introduced in Ref.~\cite{hara2020current}. We induce an asymmetry between the layers by slight deviations of the effective parameters. Due to this asymmetry  and the interlayer interaction, the motion of the electrons can be regarded as closed loops of an electrical current that allow for an in-plane OM\@. Although the Rashba 2DEG has been the first system for which the SEE has been predicted, its OEE has not been discussed yet, particularly not within the modern theory of OM\@. By extending this paradigm Edelstein system to two coupled layers and applying the modern theory of OM\@, we examine the OEE and the SEE concerning their dependence on the model parameters, and we reveal the role of layer localization of the electronic states.

 This Paper is organized as follows. In Section \ref{sec:math_back}, we set the expressions and overall framework for the spin and orbital contributions to the current-induced magnetization using a semiclassical Boltzmann approach in a quasi-two-dimensional system. In Section \ref{sec:Model}, we calculate the spin and orbital moments for a bilayer system with Rashba interaction. In Section \ref{sec:Discussion}, we discuss the spin and orbital Edelstein effects, their dependence on the model parameters, and real materials candidates hosting these double Rashba states. Finally, we conclude in Section \ref{sec:Conclusions}.

\section{\label{sec:math_back} Current-induced spin and orbital magnetization in a 2D electron gas}

At zero temperature, the magnetic moment per unit cell $\vec{m}$ in terms of spin and orbital contributions is given by 
\begin{equation}
    \vec{m} = - \frac{\mu_{\mathrm{B}}}{\hbar} \frac{A_0}{A_s} \sum_{n \vec{k}} f_{n \vec{k}}  ( g_s \sk + g_l \lk ),  
\end{equation}
where $A_0$ is the area of the unit cell, $A_s$ is the area of the entire system. $\mu_{\mathrm{B}}$ is the Bohr magneton, $\hbar$ is the reduced Planck constant. $f_{n\bk}$ is the non-equilibrium distribution function. $g_{s/l}$ are the spin and orbital $g$-factors, respectively. $\sk$ and $\lk$ are the expectation values of the spin and orbital angular momentum, respectively, and $n$ and $\bk$ indicate the band index and crystal momentum. 

Solving the linearized Boltzmann equation within constant relaxation time approximation, the distribution function in the presence of an external electric field $\vec{E}$ is $f_{n\bk} = f^0_{n\bk} + e\tau (\partial f/\partial \varepsilon)|_{\varepsilon = \varepsilon_{n\bk}} \vec{v}_{n \bk} \cdot \vec{E}$, where $f^0_{n\bk}$ is the Fermi-Dirac distribution function, $\varepsilon$ is the energy, $\vec{v}_{n \bk}$ is the group velocity, $e$ is the absolute value of the electron's charge, and $\tau$ is the constant relaxation time. 

The expectation value of the spin moment is 
\begin{equation}
    \sk = \braket{ \vec{\Psi}_{n \bk}| \hat{vec{s}} | \vec{\Psi}_{n \bk}},
    \label{eq:sk}
\end{equation}
where $\hat{\vec{s}}$ is the spin operator and $\ket{\vec{\Psi}_{n \bk}}$ is an eigenstate of the Hamiltonian. Within the modern theory of orbital magnetization \cite{thonhauser2011theory}, the expectation value of the orbital moment is defined as \cite{xiao2005berry, thonhauser2005orbital}
\begin{equation}
    \lk = \frac{ie}{2 \mu_{\mathrm{B}} g_l} \braket {\frac{\partial u_{n \bk}}{\partial \bk} | \times (\varepsilon_{\bk} - H_{0}) | \frac{\partial u_{n \bk}}{ \partial \bk} }
    \label{eq:lk}
\end{equation}
where $\varepsilon_{\bk}$ is the band energy, $H_0$ is the Hamiltonian of the system, and $|u_{n\bk} \rangle$ is the lattice-periodic part of the Bloch function. The derivative of the eigenvectors in Eq.~\eqref{eq:lk} is avoided in
\begin{equation}
    \lk = \frac{ie}{2\mu_{\mathrm{B}} g_l}  \sum_{m (\neq n)} \frac{\langle u_{n\bk} | \frac{\partial H_{0}}{\partial \bk} | u_{m\bk} \rangle \times \langle u_{m\bk} | \frac{\partial H_{0}}{\partial \bk} | u_{n\bk} \rangle}{\varepsilon_{n\bk} - \varepsilon_{m\bk}} 
    \label{eq:lk_murakami}
\end{equation}
($n$ and $m$ band indices) which does not yield all components of the OAM in 2D systems since the out-of-plane component of $\bk$ is not defined. This problem is avoided by replacing 
\begin{equation}
    \langle u_{n\bk} | \frac{\partial H_0}{\partial k_z} | u_{m\bk} \rangle = i(\varepsilon_{n\bk} - \varepsilon_{m\bk}) \langle u_{n\bk} | \hat{z} | u_{m\bk} \rangle \ ,
\end{equation}  
\noindent as suggested in Ref.~\cite{hara2020current}. Here, the system is assumed finite in the $z$-direction. In the following, $\bk$ is a 2D vector, and $\hat{z}$ is the out-of-plane component of the position operator. 

Finally, we define the Edelstein susceptibility tensor in the linear-response regime as \cite{johansson2021spin}
\begin{equation}
    \vec{m} = (\chi^s + \chi^l) \vec{E} = \chi \vec{E},
\end{equation}
where $\chi^s$, $\chi^l$ and $\chi$ are the spin ($s$), orbital ($l$) and total Edelstein susceptibilities, respectively; $\vec{E}$ is the applied electric field.

\section{\label{sec:Model}Model}
We consider a semi-infinite system with two Rashba layers at its surface (or interface to a substrate) labeled $A$ and $B$. Each layer is described by a 2D Rashba Hamiltonian and they are coupled with a spin-independent interaction. The corresponding Hamiltonian of the two-layer system is
\begin{equation}
\label{eq:model_hamiltonian}
    H = \begin{bmatrix}
H_A & T \\
T & H_B 
\end{bmatrix}
\end{equation} \ ,
where  
\begin{equation}
    H_{l} = \frac{\hbar^2 k^2}{2m_{l}} + \alpha_{l} (\unVec{z} \times \bk) \cdot \vec{\sigma}, \quad l = A, B,
    \label{eq:general_hamil}
\end{equation}
are the Rashba Hamiltonians \cite{rashba1960properties, bychkov1984properties, bychkov1984oscillatory}, with $m_{l}$ and $\alpha_{l}$ being the effective mass and Rashba parameter of layer $l = A, B$, respectively. $\unVec{z}$ is the unit vector along the surface normal, $\vec{\sigma} = (\sigma_x, \sigma_y, \sigma_z)$ are the Pauli matrices, so the spin operator in Eq. \ref{eq:sk} is $\hat{\vec{s}} = \frac{\hbar}{2} \mathbb{I}_{2\mathrm{x}2} \otimes \vec{\sigma} $. The interaction between the layers is modeled via the hopping matrix $T = t \mathbb{I}_{2\mathrm{x}2}$, with $t$ the interlayer hopping. The out-of-plane operator is defined as $\hat{z} = (c/2) \mathrm{diag}(1, -1) \otimes \mathbb{I}_{2\mathrm{x}2}$, with $c$ the distance between the layers.

\begin{figure*}
\includegraphics[width = \textwidth]{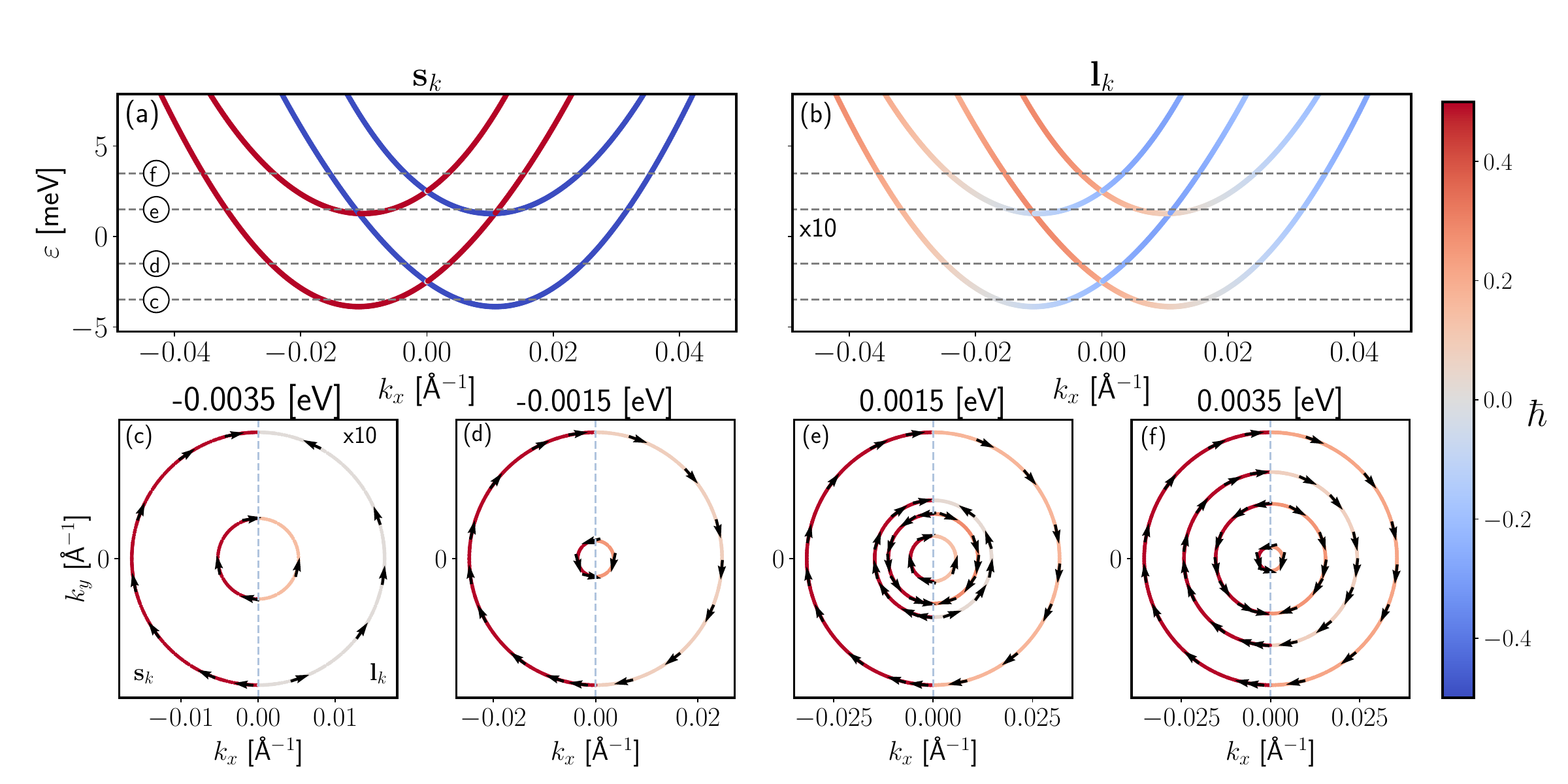}
\caption{(a)-(b) Band structure of the Rashba bilayer model along the $k_x$ axis. In color, the $y$ components of the expectation value of the spin (a) and orbital (b) moments (in units of $\hbar$). (c)-(f) Expectation values of the spin (left part of each figure) and orbital moments (right) at selected iso-energy contours, corresponding to the energies indicated by horizontal lines in (a) and (b). The color indicates the magnitude of the respective moment. The orbital moment is amplified by a factor of $10$. Here, the parameters $\alpha_A =  2 \alpha_B =  \SI{0.33}{\electronvolt \text{\AA}}$, $m_A = 2 m_B /3 = 0.27 m_e$, and $t = \SI{2.5}{m \electronvolt}$ are chosen. The absolute value of the orbital moment exhibits a clear dependence on the magnitude of $\bk$, whereas the spin moment is constant with respect to $|k|$.}
\label{fig:polarization_fermi_lines}
\end{figure*}

For the following calculations, the parameters for layer $A$ are taken from the Rashba surface states of Au(111) \cite{LaShell1996,Cercellier2006} for a reasonable order of magnitude, whereas for layer $B$, we use arbitrary ratios of $\alpha_A/\alpha_B$ and $m_A/m_B$, to introduce an asymmetry between the layers. The band structure shows two pairs of Rashba-type bands (Fig.~\ref{fig:polarization_fermi_lines}), split by $2 t$ at $\bk = 0$. The magnitude of the spin and orbital moments is constant along iso-energy lines, with their orientation locked perpendicular to $\bk$. The spin moment presents a $k$-independent magnitude (see color in Fig. \ref{fig:polarization_fermi_lines}a) with a fixed sense of rotation per band. The texture of the orbital moment shows a more complex $k$-dependent magnitude and orientation (see color in Fig. \ref{fig:polarization_fermi_lines}b).

The Rashba parameter of a layer can be associated with a potential gradient perpendicular to the interface,
\begin{align}
\alpha_R \propto \int |\Phi(z)|^2 \frac{\partial V(z)}{\partial z} \,\mathrm{d}^{3}r,
\end{align}
with $\Phi(z)$ the $z$-dependent part of an eigenstate \cite{Simon2010}, but has been shown to be affected by other factors as well, e.g., by an in-plane potential gradient \cite{Ast2007, premper2007spin}. Therefore, by having different Rashba parameters per layer, we can simulate a layer-dependent interface potential gradient, whereas to study a heterostructure, different effective masses and Rashba parameter are needed.

The Hamiltonian introduced in Eq.~\eqref{eq:general_hamil} is an effective model simulating two coupled Rashba layers, which can be employed to approximate realistic band structures around distinct points. It is based on several approximations, which will be summarized in the following. In realistic materials exhibiting deviations from this idealized model, we expect quantitatively modified results for the SEE and OEE. However, we expect qualitatively similar results for the energy window around the bilayer-Rashba-like features as long as the main contribution to the EE stems from these states. Interactions with states from other layers are neglected. Further, we assume spin-independent coupling between the two layers. The model Hamiltonian~\eqref{eq:model_hamiltonian} is not restricted to any particular orbital basis. Hence, the OEE cannot be calculated within the ACA but only by employing the modern theory of OM. Our calculations do not consider further influences of disorder, like a $\vec k$-dependent relaxation time, scattering-in terms, or lifetime broadening of the states. Contributions from edge states are also not considered.

The diagonalization of Eq.~\eqref{eq:model_hamiltonian} involves the solution of a fourth-degree equation that cannot be solved analytically for arbitrary parameters. The usual solution methods to the fourth-degree equation lead to either $k$-dependent conditions for the effective parameters ($\alpha_{A/B}(k)$, $m_{A/B}(k)$), which are out of the scope of our studies, or particular parameter combinations. In the following, we will focus on two of these particular cases: firstly, equal effective mass but different Rashba parameters ($m_A = m_B$, $\alpha_A \neq \alpha_B$), and secondly, different effective masses but equal Rashba parameters ($m_A \neq m_B$, $\alpha_A = \alpha_B$).

\subsubsection{Equal effective masses}
Assuming $m_A = m_B \equiv m$, the dispersion relation yields
\begin{equation}
    \varepsilon^{n_1, n_2} (\bk) = \frac{\hbar^2 k^2}{2m} + \frac{n_1}{2} |\alpha_{+}| k + \frac{n_2}{2}\sqrt{\alpha_{-}^2 k^2 + 4t^2}
\end{equation}
with $n_1, n_2 = \pm 1$ and $\alpha_{\pm} = \alpha_A \pm \alpha_B$. $n_1$ indicates the shape of the band, either a V-shape for the inner band ($n_1 = 1$) or a W-shape for the outer band ($n_1 = -1$), similar to a monolayer Rashba system. 

The expectation value of the spin moment reads
\begin{equation}
    \sk =  n_1 \frac{\hbar}{2} \unVec{\phi},
    \label{eq:spin}
\end{equation}

\begin{figure}
\centering
\includegraphics[width = 3.4in]{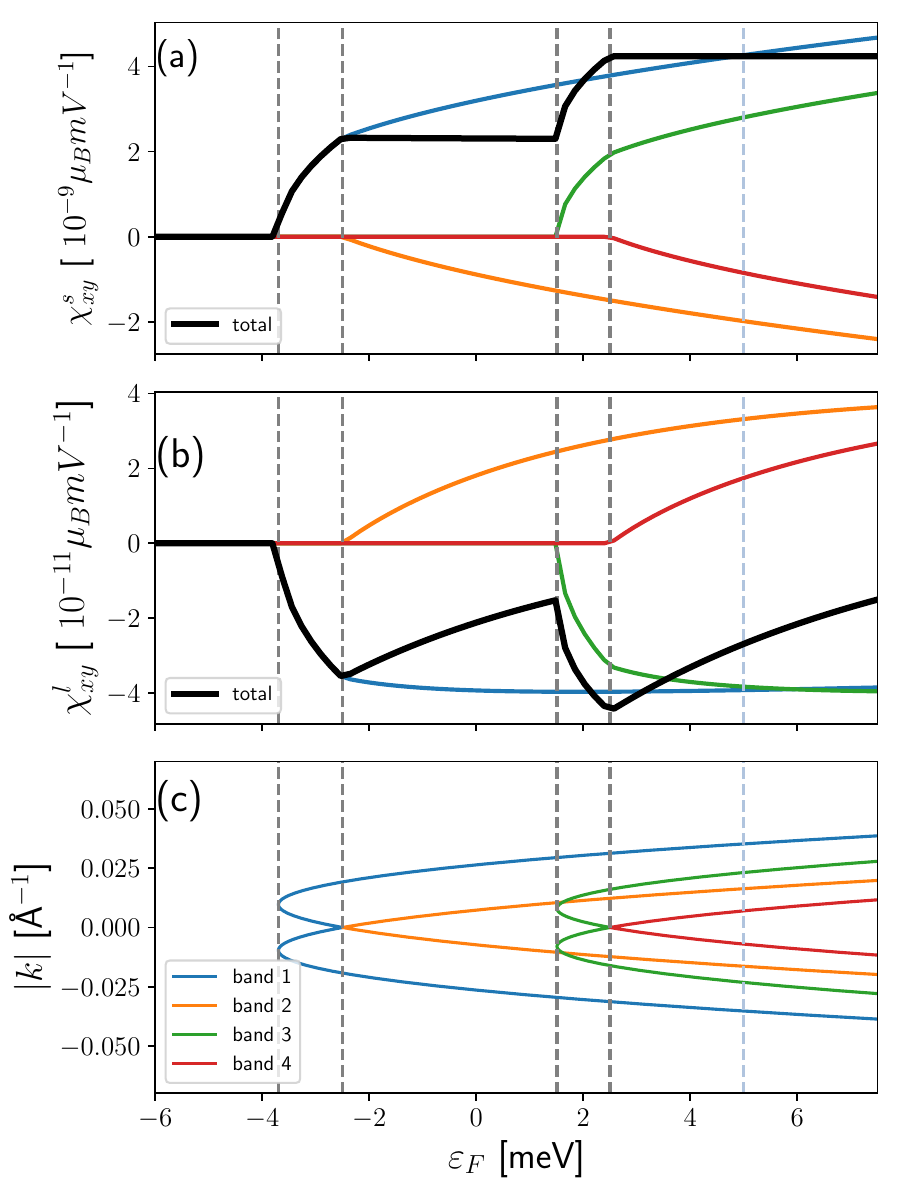}
\caption{(a-b) Total and band-resolved spin and orbital Edelstein effect and (c) corresponding band structure. The calculations where performed for the parameters $\alpha_A = 2\alpha_B = 0.33\si{\electronvolt \text{\AA}}$, $m_A = m_B = 0.27 m_e$, with $m_e$ the mass of the electron, interlayer hopping $t = 2.5 \si{m\electronvolt}$, $A_0 = 10 \si{\text{\AA}}$, and $c = 2 \si{\text{\AA}}$. The OEE has an opposite sign as the SEE, which can be seen in Eq. \eqref{eq:lk_same_mass}. The gray vertical lines indicate the energy for the bottom of each band, and the light blue vertical line highlights $\varepsilon_{\mathrm{F}} = \SI{5}{\milli\electronvolt}$ (see discussion of Figs.~\ref{fig:alpha} and \ref{fig:masses}).}
\label{fig:same_mass}
\end{figure}

with $\unVec{\phi}$ the azimuthal unitary vector in cylindrical coordinates, therefore, the absolute value of the spin moment is constant (Eq.~\eqref{eq:spin} includes a factor of $\alpha_+ / |\alpha_+|$, which is neglected here since we consider positive values of the Rashba parameters). The orientation of the expectation value of the spin moment depends on the azimuth of $\bk$ and the band shape (W or V), as for the monolayer Rashba system \cite{rashba1960properties, bychkov1984properties, bychkov1984oscillatory}; see Fig.~\ref{fig:same_mass}.

The expectation value of the orbital moment
\begin{equation}
    \lk = - n_1 \frac{e c t^2 \alpha_{-}}{\mu_\mathrm{B} g_l ( \alpha_{-}^2 k^2 + 4 t^2)} \unVec{\phi}
    \label{eq:lk_same_mass}
\end{equation}
decays with the magnitude of $k$ and, as for $\sk$, also includes $\alpha_+ / |\alpha_+|$ for the general case. It is important to note that the band's shape (W or V) determines the sense of rotation for both spin and orbital moments. In addition, the orbital moment also depends on the difference in the Rashba parameters, $\alpha_{-}$, leading to zero orbital moments for a system of two equivalent layers. 

 \subsubsection{Equal Rashba parameters}
The dispersion relation for a general combination of effective masses but equal Rashba parameters ($\alpha_A = \alpha_B \equiv \alpha$) is given by
\begin{equation}
    \varepsilon^{n1, n2} (\bk) = \frac{\hbar^2 k^2}{4} M_{+} + n_1 |\alpha| k + n_2 \sqrt{\frac{h^4 k^4 }{16} M_{-}^2 + t^2}, 
\end{equation}
with $M_{\pm} = \frac{1}{m_A} \pm \frac{1}{m_B}$. The spin expectation value is identical to that in the former case, Eq. \eqref{eq:spin}, but the orbital moment
\begin{equation}
    \lk = - \frac{ec\hbar^2 t^2 M_{-} k}{4 \mu_{\mathrm{B}} g_l \left(\frac{\hbar^4 k^4}{16} M^2_{-} + t^2 \right)} \unVec{\phi}
    \label{eq:lk_same_rashba}
\end{equation}
depends now on $k$ and does depend neither on the band index $n_1$ nor on the sign of $\alpha$.

\begin{figure}
\centering
\includegraphics[width = 3.4in]{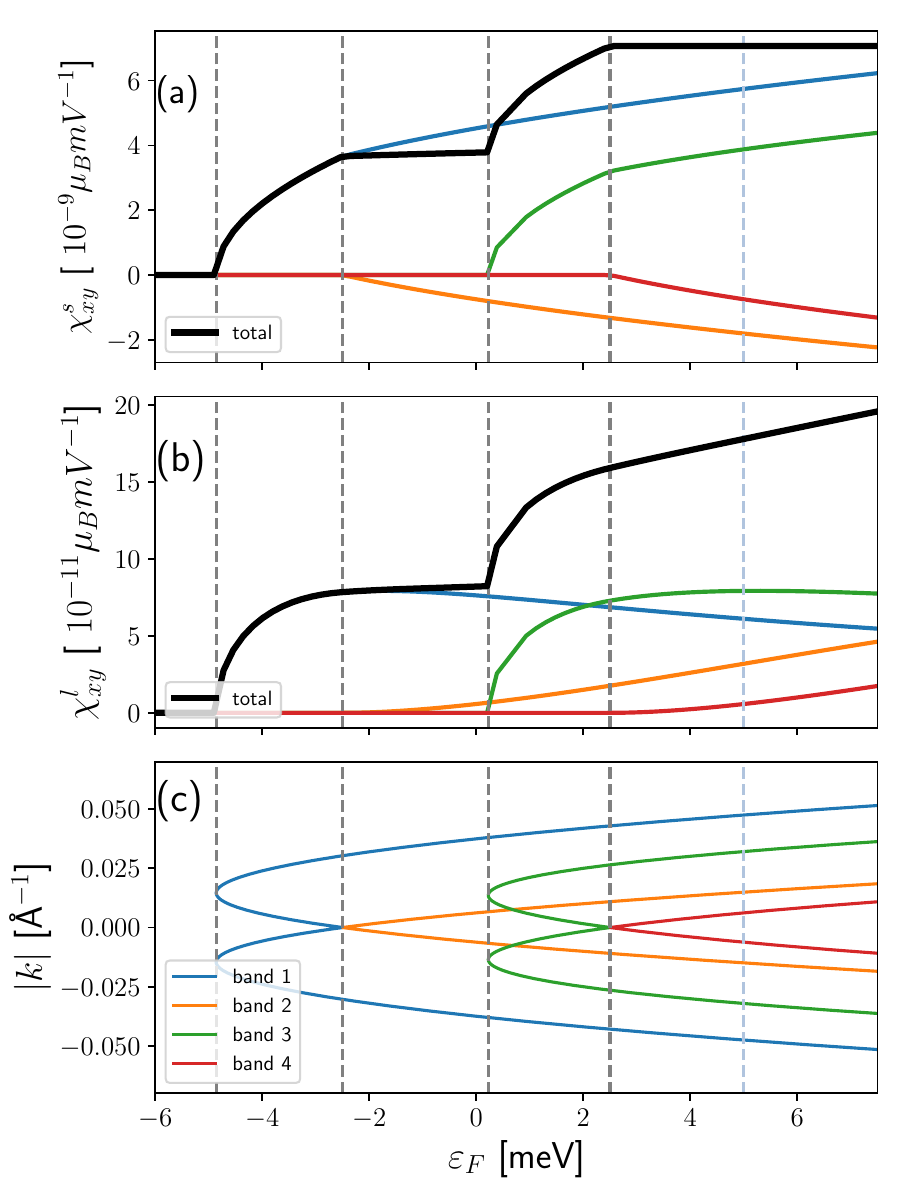}
\caption{(a-b) Total and band-resolved spin and orbital Edelstein effect and (c) corresponding band structure. The calculations where performed for $\alpha_A = \alpha_B = 0.33 \si{\electronvolt \text{\AA}}$,  $m_A = (2/3) m_B = 0.27 m_e$, interlayer hopping $t = 2.5 \si{m\electronvolt}$, $A_0 = 10 \si{\text{\AA}}$, and $c = 2 \si{\text{\AA}}$. The gray vertical lines indicates the energy for the bottom of each band, and the light blue vertical line highlights $\varepsilon_{\mathrm{F}} = \SI{5}{\milli\electronvolt}$ (see discussion of Figs.~\ref{fig:alpha} and \ref{fig:masses}).}  \label{fig:same_rashba}
\end{figure}

\section{\label{sec:Discussion} Results and discussion}
Due to the symmetries of the system introduced in Sec.~\ref{sec:Model}, particularly rotational and mirror symmetries, the only nonzero tensor elements of the Edelstein susceptibility are $\chi^{s/l}_{xy} = -\chi^{s/l}_{yx}$.

The spin and orbital moments discussed above (Eqs.~\eqref{eq:spin}, \eqref{eq:lk_same_mass}, and \eqref{eq:lk_same_rashba}), as well as the specific band structure, lead to the characteristic shape of the energy-dependent Edelstein susceptibilities shown in Figs.~\ref{fig:same_mass} and \ref{fig:same_rashba}. First, the Edelstein effect in a system with equal effective masses in both layers, shown in Fig.~\ref{fig:same_mass}, is discussed. Increasing $\varepsilon_{\mathrm{F}}$, starting from the band edge of the lowest "W"-shaped band, increases the absolute value of both the spin Edelstein effect ($\chi_{xy}^s$) and the orbital Edelstein effect ($\chi_{xy}^l$) due to the increasing number of states contributing to transport. The opposite signs originate from the opposite orientation of spin and orbital moments. When the second, "V"-shaped band is occupied, $\chi_{xy}^s$ approximately saturates due to partial compensation of the spin Edelstein effect originating from both bands, like in a monolayer Rashba system. Recall that both "W" and "V" shaped bands have spin textures with opposite senses of rotation and contribute oppositely to the SEE. Such partial compensation is also visible in the OEE signal (b). However, no saturation is visible here due to the $k$-dependence of the absolute value of the orbital moments, see Eq.\eqref{eq:lk_same_mass}, leading to a decay of the orbital susceptibility for energies between the second and third band edge. This energy-dependence is repeated qualitatively when the third and fourth bands are occupied.

In a bilayer system with equal Rashba parameters in both layers (Fig.~\ref{fig:same_rashba}), the energy-dependent SEE qualitatively behaves as in the previously discussed case of equal masses. However, the OEE exhibits qualitatively different behavior due to the band-independent sense of rotation of the orbital moments (Eq.~\eqref{eq:lk_same_rashba}). Here, spin and orbital moment are not aligned oppositely. Hence the signs of $\chi_{xy}^s$ and $\chi_{xy}^l$ are equal in the whole energy range. Further, the contributions of the "W" and "V" shaped bands do not compensate, but add up due to the equal sense of rotation of the orbital textures.

As shown in  Figs. \ref{fig:same_mass} and \ref{fig:same_rashba}, the SEE is larger than the OEE. This can be understood from  the lack of a specific orbital character of the bands for the general effective Rashba model (therefore no well-defined orbital angular momentum operator in the atom-centered approximation), avoided crossings (essential for Berry curvature-like expressions such as Eq.~\eqref{eq:lk_murakami}), the SOC leading to a pronounced SEE, and the low penetration length of the 2DEG restricting the orbital motion within the two layers.

\subsection{\label{subsec:Param_dep} Parameter dependence}

In the previous section, we have shown the results for a particular combination of the effective parameters, in which, we take the values of Au$(111)$ \cite{LaShell1996,Cercellier2006} for the first layer ($\alpha_A$, $m_A$) and different but comparable values for the second layer ($\alpha_B$, $m_B$). We have found no significant changes for other parameter combinations aside from different but qualitatively similar band structures and energy-dependent Edelstein signals. As shown in Eqs. \eqref{eq:lk_same_mass} and \eqref{eq:lk_same_rashba}, the differences between the parameters, $\alpha_-$ and $M_-$, control the size of the orbital moments. Therefore, by changing the values of the second layer's parameters, the magnitude as well as the sign of the OEE can be controlled. 

\begin{figure}
\centering
\includegraphics[width = 3.4in]{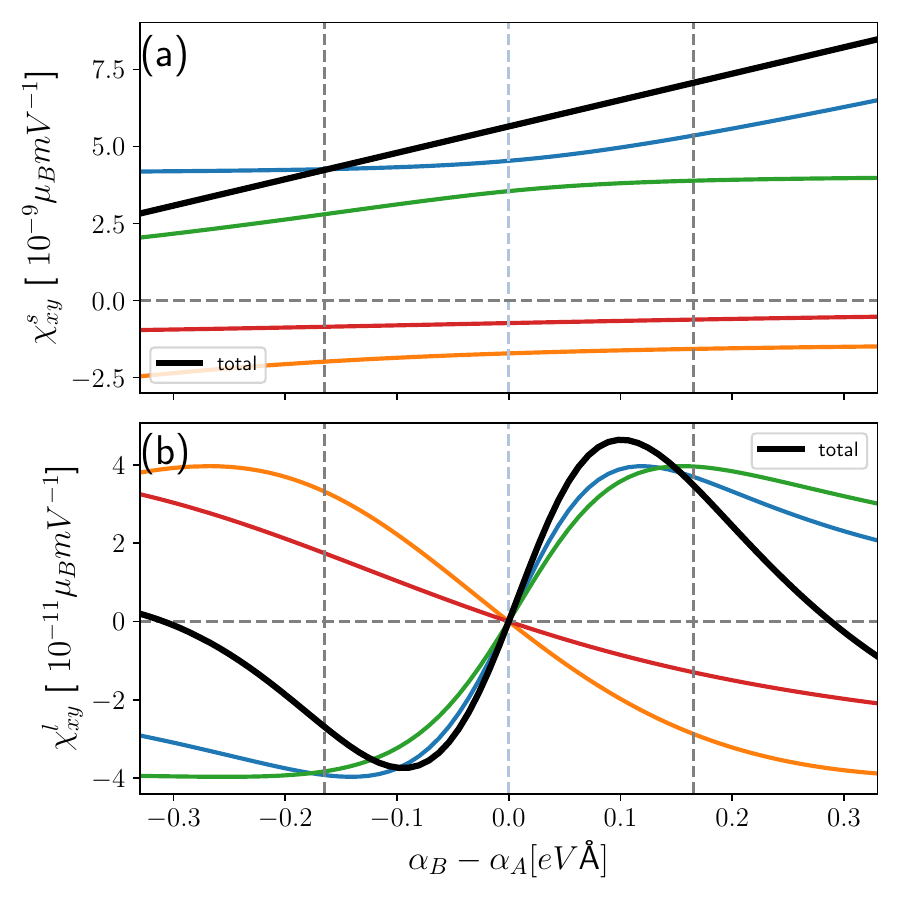}
\caption{Spin (a), orbital (b) Edelstein susceptibility for different values of $\alpha_B - \alpha_A$ and equal effective masses $m_A = m_B$. The calculations were performed for $\varepsilon_\text F= 5 \si{\milli\electronvolt}$ and the same parameters as in Fig.~\ref{fig:same_mass}, with $\alpha_A$ constant.}
\label{fig:alpha}
\end{figure}

\begin{figure}
\centering
\includegraphics[width = 3.4in]{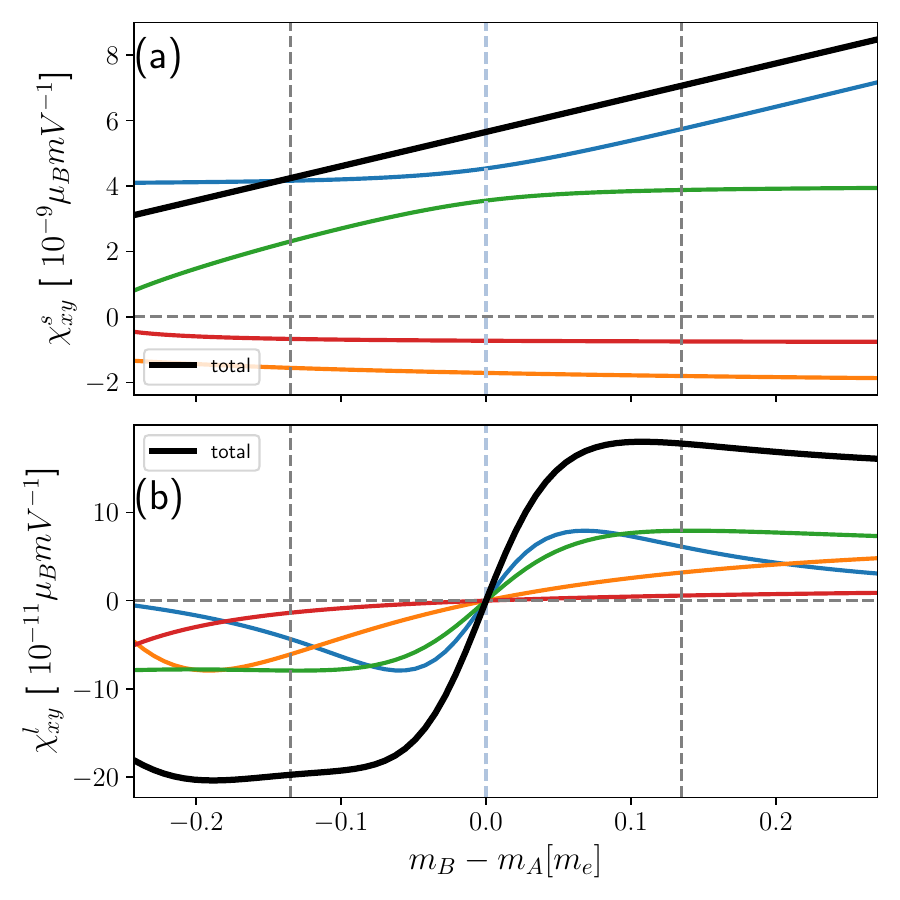}
\caption{Spin (a), orbital (b) Edelstein susceptibility for different values of $m_B-m_A$ and equal Rashba parameters $\alpha_A = \alpha_B$. The calculations where performed for $\varepsilon_{\mathrm{F}}= 5 \si{\milli\electronvolt}$ and the same parameters as in Fig.~\ref{fig:same_rashba}, with $m_A$ constant.}
\label{fig:masses}
\end{figure}

Figures \ref{fig:alpha} and \ref{fig:masses} show the spin and orbital Edelstein susceptibilities as a function of the difference between the Rashba parameters ($\alpha_B - \alpha_A$) and the effective masses ($m_B - m_A$), respectively. Both calculations are performed for the same parameters as in Figs.~\ref{fig:same_mass} and \ref{fig:same_rashba} at fixed Fermi energy ( $\varepsilon_{\mathrm{F}} = \SI{5}{\milli\electronvolt}$), changing the value of the corresponding parameter on layer $B$, $\alpha_B$ and $m_B$, respectively. As shown in Figs.~\ref{fig:alpha}a and \ref{fig:masses}a, the SEE is enhanced by increasing either $\alpha$ or $m$ in one of the layers. The approximately constant increase of the SEE is related to an increasing size of the Fermi lines. However, the SEE can present the opposite sign for a system with a negative sum of the Rashba parameters, i.e., $\alpha_A + \alpha_B <0$, although that configuration is not studied in the present work.

Figure~\ref{fig:alpha}b shows that the sign of the OEE is controlled by the difference of the Rashba parameters, leading to a sign change for the case of equivalent layers ($m_A = m_B$ and $\alpha_A = \alpha_B$), related to the symmetry of the system discussed in the following subsection, and a two more parameter-dependent sign change around $\alpha_B - \alpha_A \approx \pm 0.3 \si{\electronvolt \angstrom}$. The OEE for equal Rashba parameters, shown in Fig. \ref{fig:masses}b, only shows a sign change for the case of equivalent layers. The sign change of $\alpha_B - \alpha_A$ and $m_B-m_A$, respectively, means a reversed orbital moment at each $\bk$ point, hence a reversed sense of rotation of the orbital moment along the iso-energy lines and a sign change of the OEE. One important difference between both cases, equal $m$ and equal $\alpha$, is the band independence of Eq.~\eqref{eq:lk_same_rashba}, since for the equal Rashba parameters case, all bands exhibit the same sense of rotation of the orbital moments, contributing with the same sign to the OEE in Fig.~\ref{fig:masses}b. 


In contrast to the SEE, which exists only due to SOC, orbital effects do not require SOC, which has been confirmed by studies on various systems without SOC, such as free-electron systems and chiral structures~\cite{Go2018, yoda2018orbital} by introducing different types of asymmetries. In Fig.~\ref{fig:alpha}b, the OEE at $\alpha_B =0$ ($\alpha_B - \alpha_A = -0.33\si{\electronvolt \text{\AA}}$ in the figure), although small, is not zero, showing that the OEE can occur in the absence of Rashba splitting in one of the layers. In the Appendix~\ref{App:A}, we  further discuss a bilayer free-electron system without Rashba SOC, but with asymmetric effective masses, and show that here a finite OEE occurs, whereas the SEE vanishes.

\subsection{\label{subsec:Layer_dep} Layer dependence}

\begin{figure*}
\centering
\includegraphics[width = 7in]{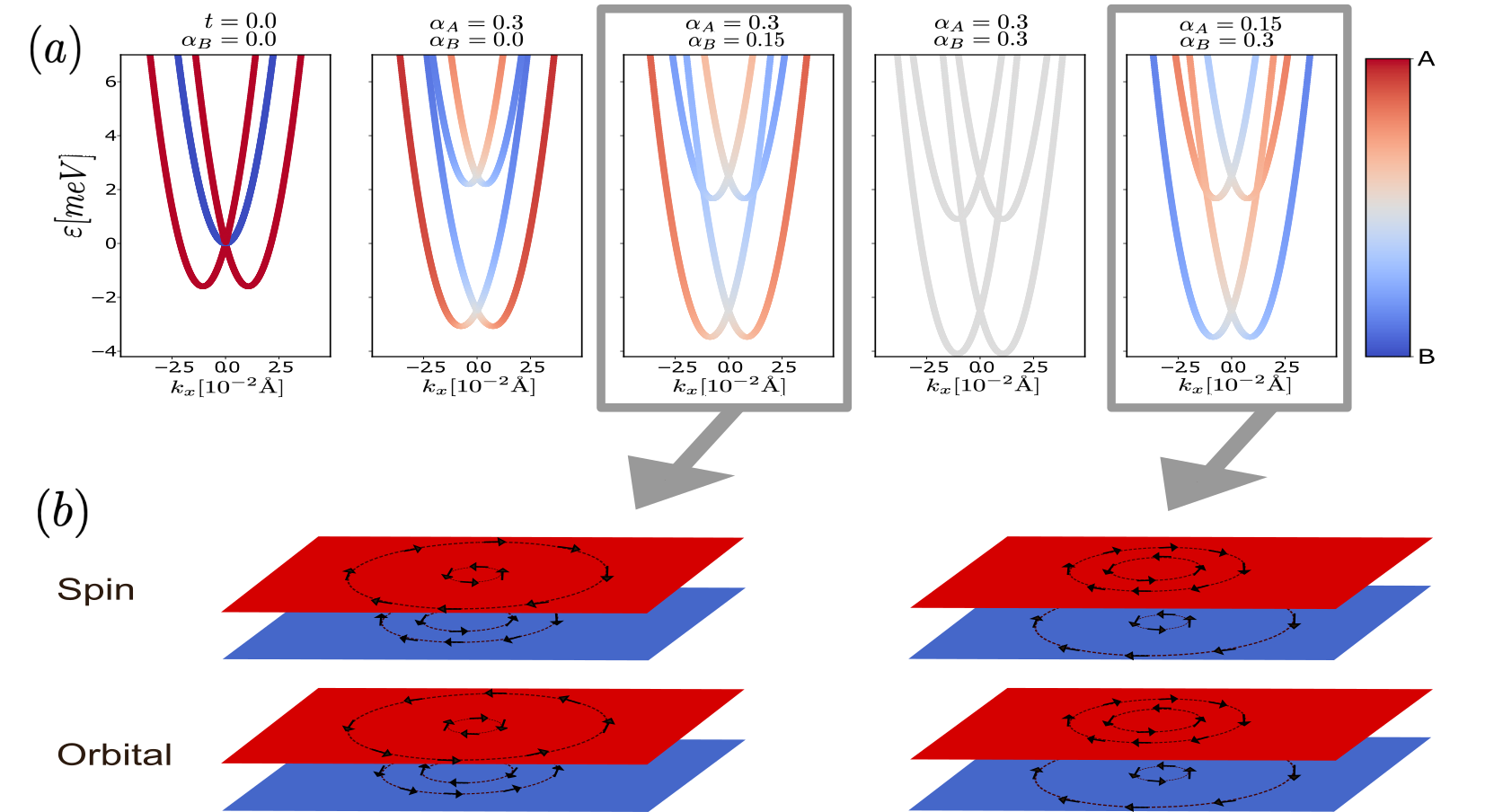}
\caption{(a) Layer projection of the eigenstates for different values of the Rashba parameters in the system, performed for the same parameters as Fig.~\ref{fig:same_mass}, except for those indicated in the Figure's titles. (b) Sketch of the iso-energy lines with spin and orbital textures, projected to the corresponding layers for $\varepsilon_{\mathrm{F}}= ~ 5 \si{\milli\electronvolt}$.}
\label{fig:layers}
\end{figure*}

From Eqs. \eqref{eq:lk_same_mass} and \eqref{eq:lk_same_rashba}, it is clear that a sign change of $\alpha_-$ and $M_-$, respectively, induces a sign change of the $\bk$-dependent orbital moment per band, $\lk$. A physical interpretation of the origin of the sign change in OEE can be obtained by analyzing the localization of the states per layer. In contrast to the spin moment, the OEE is tied to the out-of-plane position of the layers ($z$), giving relevance to the spatial order of the layers relative to each other. Figure \ref{fig:layers}a shows the projection of the eigenstates to the layers of a system with equal effective masses but different Rashba parameters. First, when $t=0$ and $\alpha_B = 0$, the states are fully localized in a degenerate free-electron band for layer $B$ and a simple Rashba band structure for layer $A$. At $\bk =0$ the state is four-fold degenerate. However, when we include interlayer hopping ($t \neq 0$), the degeneracies are lifted. At $\bk =0$, we observe two two-fold degenerate bands with a band gap of $2t$. The states are weakly localized in both layers around $\bk =0$. For this case, even with $\alpha_B = 0$, the states localized in layer B show an energy-splitting close to $\bk=0$. Nevertheless, this band-splitting becomes negligible for higher energies ($\varepsilon_{\mathrm{F}} \gg t$). In contrast, the states localized in layer $A$ show a Rashba-like structure with the same band gap of $2t$ at $\bk=0$. Therefore, the interlayer hopping induces Rashba interaction from layer $A$ into layer $B$, even when $\alpha_B = 0$.

For the case of equivalent layers, each eigenstate is equally localized in both layers, which can be interpreted as a total compensation of the layer contributions to the orbital moment, see Eqs. \eqref{eq:lk_same_mass} and \eqref{eq:lk_same_rashba}. This compensation is better seen when we compare two configurations of the Rashba parameters. The first configuration is when the Rashba parameter of layer $A$ is larger than of layer $B$ ($ \alpha_A > \alpha_B$), while the second is the interchanged relation ($\alpha_A < \alpha_B$). Both configurations are highlighted with grey boxes in Fig. \ref{fig:layers}a. In addition, Fig. \ref{fig:layers}b sketches spin and orbital textures along iso-energy lines, projected to the layers, for these two configurations. Comparing these two cases proves helpful since the band structure is equivalent, but the localization of the eigenstates is opposite. Here, the states on the most outer and most inner bands (1 and 4) are localized in the layer with the larger Rashba parameter, while the states of the middle bands (2 and 3) are localized in the layer with the smaller Rashba parameter. This interchange of the localization does not affect the sign of the spin expectation values (see Fig.~\ref{fig:layers}b) since the spin texture is conserved. Even though the contribution per layer changes when the localization of the eigenstates is reversed, the total current-induced spin moment remains the same. However, for the orbital moment the texture's sense of rotation per band is changed by reversion of the eigenstates' localization, which is also evident from Eq. \eqref{eq:lk_same_mass} due to the sign change of $\alpha_-$.

To quantify a layer's contribution to the orbital moments, we decompose the eigenstates as $|u_{n\bk} \rangle = |A, n\bk \rangle + |B, n\bk \rangle$, with
\begin{align}
 |A, n\bk \rangle &  = \frac{1}{N} \begin{pmatrix}
 u^A_{\uparrow, n\bk} \\ u^A_{\downarrow, n\bk} \\ 0 \\ 0
 \end{pmatrix}
 \end{align}
and analogously for $|B, n\bk \rangle$, with $1/N$ the normalization factor. With this decomposition, the orbital moment ~\eqref{eq:lk_murakami} is a sum of four terms,
$\lk = \vec{l}_{n \bk}^{AA} + \vec{l}_{n \bk}^{AB} + \vec{l}_{n \bk}^{BA} + \vec{l}_{n \bk}^{BB}$, with
\begin{equation}
    \vec{l}_{n \bk}^{XY} = \frac{ie}{2 \mu_{\mathrm{B}} g_l} \sum_{m\neq n} \frac{\langle X, n\bk| \frac{\partial H}{\partial \bk}|X, m\bk \rangle \times \langle Y, m\bk| \frac{\partial H}{\partial \bk}|Y, n\bk \rangle}{\varepsilon_{n \bk} - \varepsilon_{m \bk}},
    \label{eq:orbital_separation}
\end{equation}

($X, Y = A, B$). These contributions read 

\begin{subequations}
 \begin{align}
     \lk^{AA} & = \frac{-ect^2}{2\mu_{\mathrm{B}} g_l (k^2 \alpha_{-}^2 + 4 t^2 )} \left( \frac{\hbar^2 k}{m} + n_1 \alpha_A \right) \unVec{\phi},
 \\
     \lk^{AB} & = \frac{-ect^2}{2 \mu_{\mathrm{B}} g_l ( k^2 \alpha_{-}^2 + 4 t^2 )} \frac{n_1 \alpha_{-}}{2} \unVec{\phi},
 \\
     \lk^{BB} & = \frac{ect^2}{2 \mu_{\mathrm{B}} g_l (k^2 \alpha_{-}^2 + 4 t^2 )} \left( \frac{\hbar^2}{m} + n_1 \alpha_B \right) \unVec{\phi}
 \end{align} 
\end{subequations}

for a system with $m_A = m_B$; confer Eq.~\eqref{eq:lk_same_mass}. The mixed or interlayer terms are equal ($\lk^{AB} = \lk^{BA}$), but the intralayer terms $\lk^{AA}$ and $\lk^{BB}$ have opposite sign and differ according to the respective Rashba parameters, or analogously according to the effective masses for a system with $\alpha_A = \alpha_B$. Therefore, the physical origin of the nonzero $\bk$-dependent orbital moment can be attributed to the asymmetry in the layer-wise contributions, since electrons flowing between the layers acquire an orbital motion in out-of-plane trajectories, which, in analogy to a loop of electrical current, generates an in-plane orbital moment~\cite{hara2020current}.

Applying the above decomposition to the spin moment, Eq.~\eqref{eq:spin}, shows that only the intralayer terms contribute to the SEE, both with the same sign, which is a direct consequence of the definition of the spin moment and the interactions in the system. Therefore, the addition of extra layers does not introduce new physics to the spin moment in our model. However, as shown in Ref. \cite{hara2020current}, the extension of the 2DEG to the third dimension has an important role for the magnitude  of the OEE. We discuss the effect of an extended 2DEG for the SEE and OEE in Appendix \ref{App:B}.

\subsection{Materials proposal}

 Materials showing a Rashba effect are widely used for spin-charge interconversion. Especially at oxide interfaces \cite{Ohtomo2002, Thiel2006, vaz2019mapping} and polar semiconductors \cite{chen2021prediction, Portugal2021}, 2DEGs with a thickness of several unit cells can exhibit more than one band splitting related to the Rashba effect. However, those bands are required to be energetically close to induce a sizable nonlocal contribution to the OEE\@.
 
 Polar semiconductors \cite{scanlon1959polar,eremeev2012ideal,sakano2013strongly} are suitable candidates for showing a double Rashba band structure similar to the one shown in Fig.~\ref{fig:polarization_fermi_lines}a. Al$_2$O$_3$ covered by a monolayer of a heavy metal has been reported to host similar double Rashba band structures. In this substrate, Al atoms are located at a slightly different height than the O atoms due to surface relaxation. Therefore, the monolayer of the heavy metal (Pb, Bi, Sb, and their ordered alloys \cite{chen2021prediction}) is expected to form a buckled adlayer \cite{cahangirov2009two, chen2021prediction}. 

 Recent works have suggested a sizeable orbital contribution to the Edelstein effect compared to the spin contribution for oxide interfaces. Particularly, recent publications on SrTiO$_3$ \cite{johansson2021spin} and KTaO$_3$ \cite{varotto2022direct} based interfaces have shown a significant orbital magnetization using the ACA approach. These materials present energetically close bands from different layers, hinting at a relevant nonlocal contribution from the modern theory of OM. Especially for oxide interfaces in which the 2DEG is extended to several layers~\cite{sing2009profiling, vaz2019mapping}, a double or multi-layer approach, which is crucial for the application of the modern theory of orbital magnetization, is appropriate. Other oxide-based materials reported to exhibit a significant EE are BaSnO$_3$ and ZnO \cite{Trier2021}. 

 A surface polarization due to a slight spatial displacement between the atoms at the surface is key in obtaining a double (or multiple) Rashba structure from an inhomogeneous potential gradient. Therefore, the ferroelectric Rashba semiconductors (FERSC) have been proposed for purely electrical control of the Rashba interaction, even reaching a switchable configuration \cite{daSilveira2016, Varignon2019}. Other systems with switchable Rashba SOC have been reported from the perovskite family \cite{Trier2021}. Therefore, the enhancement and reversion of the orbital contribution explored in this paper could further contribute to the overall electrical control of the total conversion efficiency.

\section{\label{sec:Conclusions}Conclusions}

This paper introduces an effective model for a bilayer system with Rashba interaction to describe the current-induced spin and orbital Edelstein effect. Because of a sizeable interlayer hopping electrons can perform out-of-plane motion which allows for an in-plane OM. Here, we see that the asymmetry of the parameters of those layers is fundamental for an orbital moment. Two cases, namely equal effective masses and equal Rashba parameters of the two layers, are discussed in detail. For any parameter combination, spin and orbital moments are locked perpendicular to the momentum. The spin expectation values are constant, but the orbital moments' absolute values decay with $k$. 

We explore the model parameter dependence of the current-induced magnetization. For constant energy, the SEE is enhanced by increasing the value of either effective mass or Rashba parameter regardless of the ratio of the corresponding parameters of both layers. However, the sign of the OEE can be tuned according to the difference between the parameters, with the OEE vanishing if the layers are equivalent. The sign change of the OEE is accompanied by a change of the layers localization of the eigenstates. Tuning the ratio $\alpha_B/\alpha_A $ (or $m_B/m_A $) from $<1$ to $>1$ and vice versa, and assuming $m_B=m_A$ ($\alpha_B= \alpha_A$), the layer localization of the individual states is reversed. For the orbital moment, the sense of rotation along an iso-energy line is also reversed, whereas the spin's sense of rotation is preserved. Considering the intra- and inter-layer contributions to the orbital moment, we find that both layers contribute oppositely to the total k-dependent orbital moment, and hence the difference between the respective parameters ($\alpha_A - \alpha_B $ and $m_A-m_B$, respectively) determines the sign of the total OEE.

 The approach expressed in this work shows that the orbital moment is relevant even for systems where the expectation value of the orbital angular momentum operator is zero within the atom-centered approximation, implying that the modern theory of orbital magnetization can be essential for the discussion of orbital transport effects.

\begin{acknowledgments}
This project has received funding from the European Union's 2020 research and innovation programme under the Marie Skłodowska-Curie grant agreement No 955671. The authors thank N.G.J. for patience throughout this work.
\end{acknowledgments}

\appendix

\section{\label{App:A} OEE without SOC}

\begin{figure}
    \centering
    \includegraphics[width = 3in]{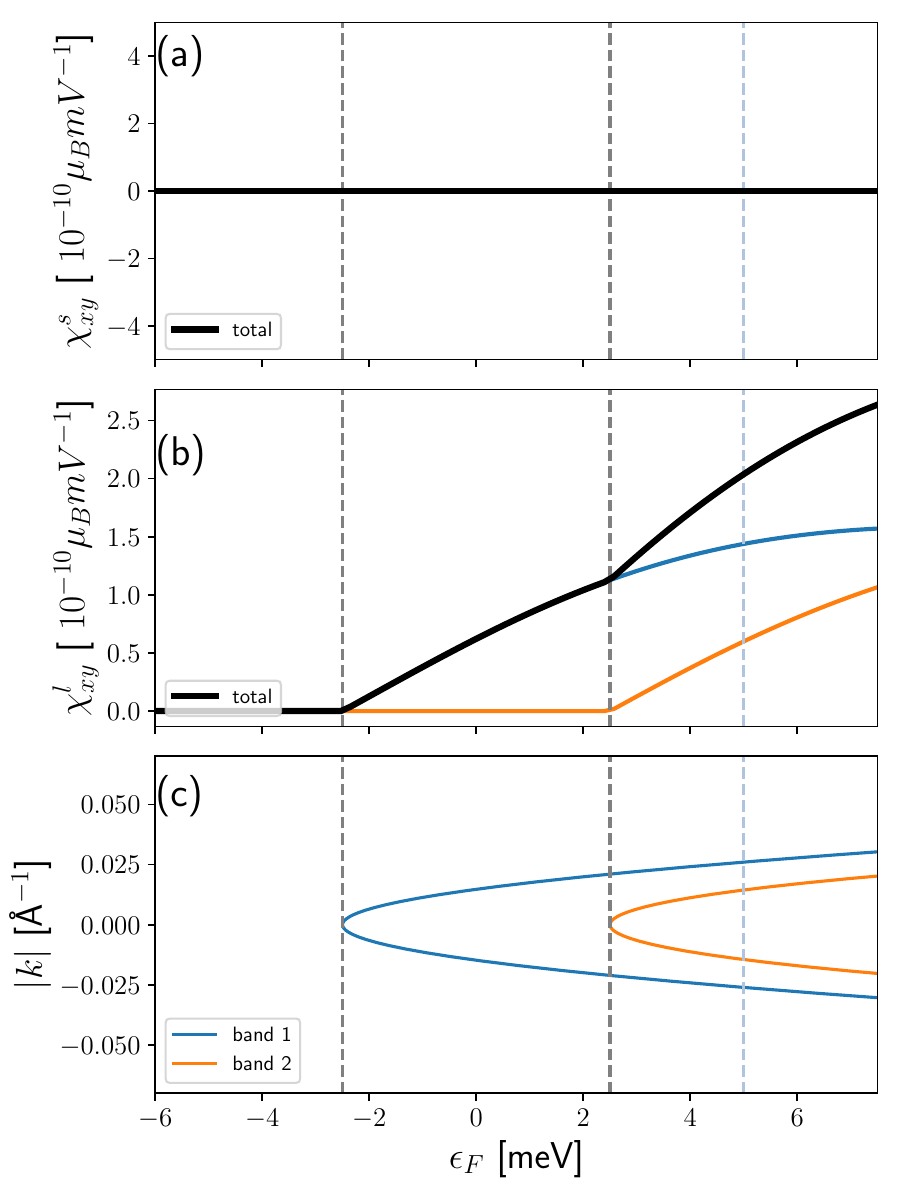}
    \caption{(a-b) Total and band-resolved spin and orbital Edelstein effect, and (c) corresponding band structure. The calculations were performed for the parameters $\alpha_A = \alpha_B = 0 \si{\electronvolt \text{\AA}}$, $m_A = (2/3) m_B = 0.27 m_e$, interlayer hopping $t = 2.5 \si{\milli\electronvolt}$, and $c = 2 \si{\text{\AA}}$.
    }
    \label{fig:noSOC}
\end{figure}

 It is well known that for the SEE, SOC is crucial, whereas it has been proven that the OEE does not require SOC \cite{Go2018, yoda2018orbital}. In our model, we study the Edelstein effect in a bilayer system with Rashba SOC. Obviously, $\lk \neq 0$ in the absence of SOC ($\alpha_A = \alpha_B=0$), as long as $m_A \neq m_B$, see Eq.~\eqref{eq:lk_same_rashba}. Hence, a nonzero OEE can be induced in a coupled bilayer free electron gas system with anisotropic effective masses. Fig.~\ref{fig:noSOC} shows the spin and orbital Edelstein effect as well as the band structure of a free electron gas bilayer ($\alpha_A = \alpha_B = 0$ and $m_A\neq m_B$). Due to the absence of SOC, the SEE is zero within the whole energy range, but the OEE is nonzero and increases with energy.

\section{\label{App:B} Extended 2DEG}

As stated in Section \ref{sec:Discussion}, one reasons why the SEE is larger than the OEE is the low extension of the 2DEG over only two layers. However, the spread of the 2DEG that determines the number of layers taken into account can be easily explored. Similarly, a semi-infinite insulator has been explored in Ref. \cite{hara2020current}.

 \begin{figure}
\centering
\includegraphics[width = 3in]{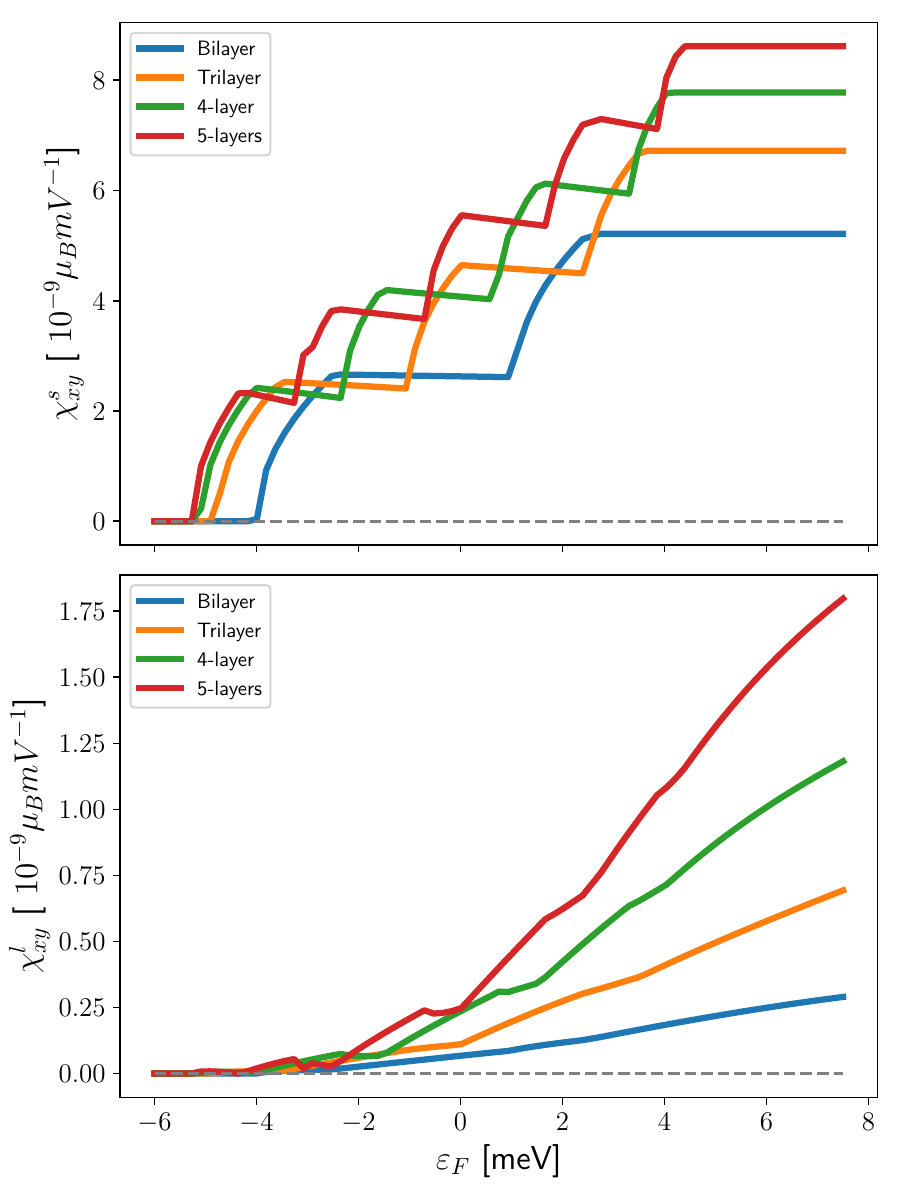}
\caption{Spin (a) and orbital (b) Edelstein susceptibilities in a quasi-2DEG consisting of two, three, four and five layers, respectively. In this configuration, the Rashba parameters decrease from the surface/interface (layer 0) to the bulk, while the effective mass increase from the surface/interface, up a maximum value for the bulk. The calculations were performed for the parameters follow Eqs.~\ref{eq:param_prog}, interlayer hopping $t = 2.5 \si{\milli\electronvolt}$ for all interlayer hoppings, and $c = 2 \si{\text{\AA}}$.
}
\label{fig:case1}
\end{figure}

 Including more layers close to the interface allows more parameter combinations. In the following, we focus on two distinct configurations. First, we add layers with decaying Rashba interaction and increasing effective mass, following a Gaussian function
 
\begin{subequations}
 \begin{align}
    \alpha_i & = \alpha_0 \exp\left({-\frac{(c i)^2}{4N}}\right),
 \\
     m_i & = m_0 \left[2-\exp\left({-\frac{(c i)^2}{4N}}\right)\right],
 \end{align} 
 \label{eq:param_prog}
\end{subequations}

where $(\alpha_0, m_0) = (\alpha_A, m_A)$ from Au(111), $c$ is the spacing between the layers, $N$ is the number of layers of the system, $i=0,1,2,3,4$ is the index of the layer. Although the layer-dependence in Eq. \eqref{eq:param_prog} is arbitrary, it is a reasonable way to extend the 2DEG into the bulk. Fig.~\ref{fig:case1} shows the spin and orbital Edelstein efficiencies for this parameter relation, simulating a 2DEG with a depth of 2 to 5 unit cells. Both  spin and orbital Edelstein effects increase due to the extra contributions from the extra layers but the orbital moment shows the stronger sensitivity towards the number of layers. Whereas the SEE is increased by a factor of $2$ comparing the bilayer and 5-layer configuration, the OEE is enhanced by a factor of $5$.

\begin{figure}
    \centering
    \includegraphics[width = 3in]{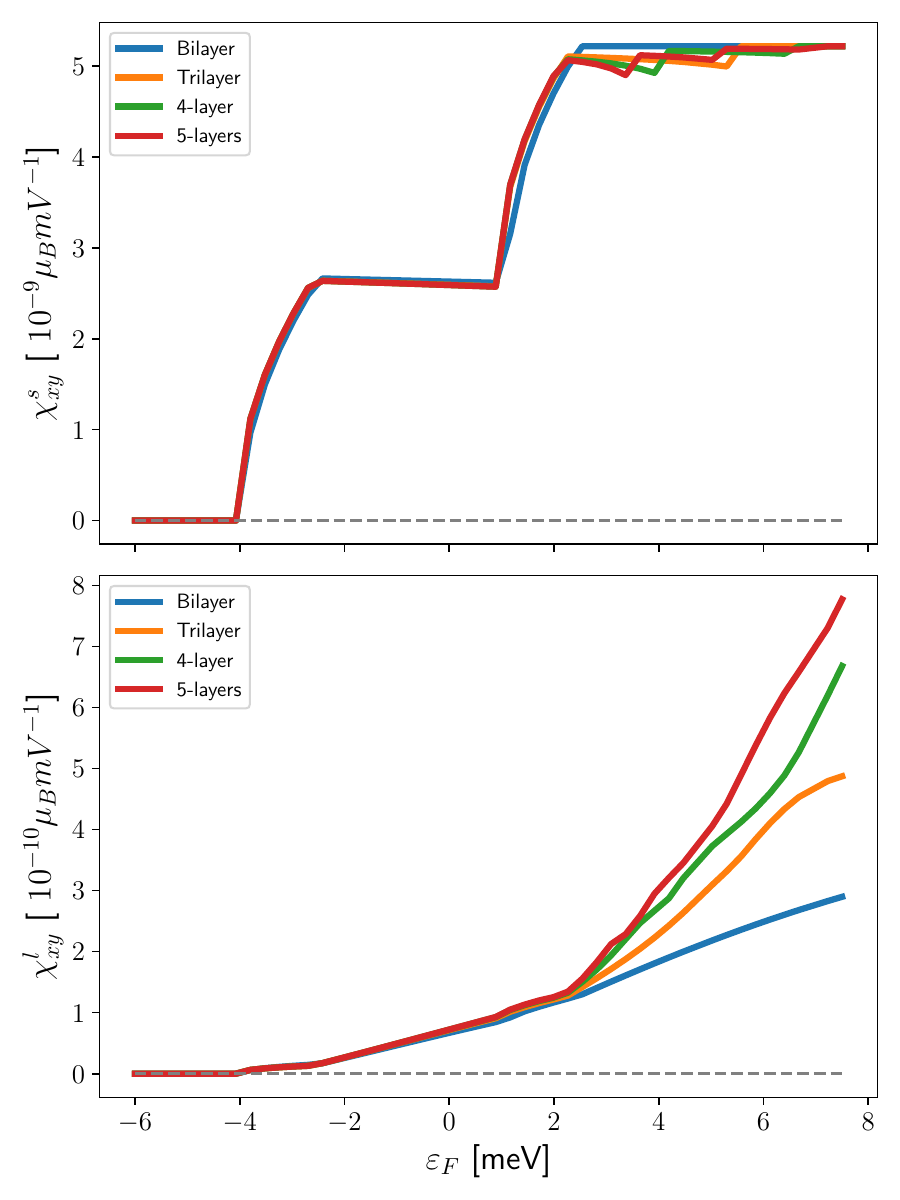}
   \caption{Spin (a) and orbital (b) Edelstein susceptibilities in a quasi-2DEG consisting of two, three, four and five layers, respectively. This configuration assumes that after the second layer ($i=1$), the potential gradient can be neglected and the following layers can be modeled  by a free electron model. The calculations were performed for the parameters following Eq.~\eqref{eq:param_prog} for the first 2 layers and $\alpha_i = 0$ and $m_i = m_1 $, with $i = (2,3,4)$ for the other layers. Different on-site energies for the extra layers are $e_i = 5 \si{\milli\electronvolt}$ for $i = (2,3,4)$, the interlayer hopping between $i=0$ and $i=1$ layers is $t = 2.5 \si{\milli\electronvolt}$ and between the other layers is $t´ = 1.25 \si{\milli\electronvolt}$, and $c = 2 \si{\text{\AA}}$.
   }
    \label{fig:case2}
\end{figure}
 
 The second configuration, for which the SEE and OEE are shown in Fig.~\ref{fig:case2}, follows Eq.~\eqref{eq:param_prog} for the first two layers ($i=0,1$), but adds  free electron-like layers for the layers $2, 3,$ and $4$ with the same effective masses per layer as the second layer ($i=1$), i.e., $\alpha_i = 0$ and $m_i = m_1 $, with $i = (2,3,4)$. For this configuration we keep the zero on-site energies for the first two layers, but for the extra layers the on-site energies are $e_i = 5 \si{m\electronvolt}$ for $i = (2,3,4)$, the interlayer hopping between the layers $0$ and $1$ is $t = 2.5 \si{m\electronvolt}$ and between the other layers is $t´ = 1.25 \si{m\electronvolt}$. The SEE presented in Fig.~\ref{fig:case2} shows almost no increase when the number of layers is enhanced, whereas the OEE increases by  a factor of $2$ at high energies, comparing the five-layer case to the two-layer case. 

The calculated Edelstein susceptibilities for both configurations show that the number of layers contributing to the quasi-2DEG has significant influence on the orbital Edelstein effect (calculated within the modern theory of orbital magnetization), whereas the spin Edelstein effect is less affected by a higher number of layers. This finding opens up new perspectives in the search for materials exhibiting a large current-induced orbital magnetization.

\typeout{} 
\bibliography{Bib}

\end{document}